\documentclass[12pt]{article}

\usepackage{scicite}
\usepackage{braket}
\usepackage{graphicx} 
\usepackage{float}
\usepackage{ragged2e}
\usepackage{subcaption}
\usepackage{tikz} 
\usepackage{amsmath}

\usepackage{times}

\newcommand{\Yb}{$^{171}\rm{Yb}^+$}

\newenvironment{sciabstract}{%
\begin{quote} \bf}
{\end{quote}}

\newcounter{lastnote}

\title{A Single-Ion Information Engine for Charging Quantum Battery}

\author
{Jialiang Zhang,$^{1\dagger}$ Pengfei Wang,$^{2\dagger}$ Wentao Chen,$^{2}$ Zhengyang Cai,$^{1}$\\ Mu Qiao,$^{1}$ Riling Li,$^{3}$ Yingye Huang,$^{1}$ Haonan Tian,$^{1}$ Henchao Tu,$^{1}$\\ Kaifeng Cui,$^{4}$ Leilei Yan,$^{4}$ Junhua Zhang,$^{5,6}$ Jingning Zhang,$^{2}$ \\Manhong Yung,$^{5,6}$ Kihwan Kim,$^{1,2,7,8\ast}$
\\
\normalsize{$^{1}$State Key Laboratory of Low Dimensional Quantum Physics, Department of Physics,  }\\
\normalsize{Tsinghua University, Beijing 100084, China}\\
\normalsize{$^{2}$Beijing Academy of Quantum Information Sciences, Beijing 100193, China}\\
\normalsize{$^{3}$Key Laboratory of System Software (Chinese Academy of Sciences)}\\
\normalsize{and State Key Laboratory of Computer Science, Institute of Software,}\\
\normalsize{Chinese Academy of Sciences, Beijing 100086, China}\\
\normalsize{$^{4}$School of Physics, Zhengzhou University, Zhengzhou 450001, China}\\
\normalsize{$^{5}$Shenzhen Institute for Quantum Science and Engineering,}\\
\normalsize{Southern University of Science and Technology, Shenzhen 518055, China}\\
\normalsize{$^{6}$International Quantum Academy, Shenzhen 518048, China}\\
\normalsize{$^{7}$Hefei National Laboratory, Hefei 230088, China}\\
\normalsize{$^{8}$Frontier Science Center for Quantum Information, Beijing 100084, China}\\
\normalsize{$^\ast$To whom correspondence should be addressed. }\\
\normalsize{E-mail:  kimkihwan@gmail.com.}\\
\normalsize{$^{\dagger}$These authors contributed equally to this work.}
}

\date{}

\begin{document} 


\baselineskip24pt


\maketitle


\begin{sciabstract}
  Information engines produce mechanical work through measurement and adaptive control. For information engines, the principal challenge lies in how to store the generated work for subsequent utilization. Here, we report an experimental demonstration where quantized mechanical motion serves as a quantum battery and gets charged in repeated cycles by a single trapped-ion information engine. This is enabled by a key technological advancement in rapid state discrimination, allowing us to suppress measurement-induced disturbances. Consequently, we were able to obtain a charging efficiency over 50\% of the theoretical limit at the optimal temperature. The experimental results substantiate that this approach can render trapped ions a promising platform for microscopic information engines with potential applications in the future upon scaling up.
\end{sciabstract}

As heat engines cyclically convert thermal energy into mechanical work \cite{landau2013statistical}, an information engine is a cyclic device that can convert the abstract concept of information into usable forms of energy. The interplay between information and thermodynamics dates back to Maxwell's demon, a hypothetical agent designed to challenge the second law of thermodynamics \cite{knott1911life}. This concept was later instantiated as a single-particle engine, known as the Szilard engine \cite{szilard1929entropieverminderung}. Recent advancements in experimental techniques have made it possible to explore information-to-work conversion in laboratory settings. Classical information engines have been demonstrated using colloidal particles, biological systems, single-electrons, and others \cite{toyabe2010experimental,roldan2014universal,koski2014experimental,parrondo2015thermodynamics,saha2021maximizing}. With the advent of quantum mechanics and the development of quantum technologies, theoretical and experimental studies of information-to-work conversion have been extended to the quantum realm \cite{scully2003extracting,kim2011quantum,park2013heat,zurek2018eliminating,myers2022quantum}. These studies aim to understand the role of quantum resources like quantum memory, coherence, and entanglement in information engines \cite{naghiloo2018information,bresque2021two,myers2022quantum} and to realize quantum information engines in various physical platforms \cite{camati2016experimental,mohammady2017quantum,cottet2017observing,chida2017power,masuyama2018information,peterson2020implementation,wang2022experimental,spiecker2023two}. However, the experimental realization of a quantum information engine that embodies the fundamental elements of cyclical operation and generates useful work, such as charging quantum batteries \cite{campaioli2024colloquium,mitchison2021charging,yan2023charging}, remains elusive.

Here, we implement the cyclic quantum information engine using a single trapped ion. Trapped ions offer exceptional control over internal and external degrees of freedom \cite{leibfried2003quantum,chen2021quantum}, making them an ideal platform. Prior works have explored single-ion heat engines in the classical regime \cite{rossnagel2016single}, verified the quantum Jarzynski equality and Landauer's principle \cite{an2015experimental,yan2018single}, studied heat engine dynamics using the internal state \cite{zhang2022dynamical}, and implemented a quantum absorption refrigerator with mechanical oscillators \cite{maslennikov2019quantum}. Recently, a quantum heat engine was developed, utilizing the internal state as the engine coupled to a mechanical oscillator as a flywheel or quantum load \cite{von2019spin,van2020single}. However, developing a cyclic information engine to charge the oscillator remained challenging due to disturbances from measuring the internal state through fluorescence \cite{hempel2013entanglement,wan2014precision,de2022error}. We suppress these measurement disturbances by developing fast state detection, enabling the successful realization of a cyclic quantum information engine capable of charging the quantum battery, the mechanical oscillator.

Our information engine is a realization of the Szilard engine in the quantum system. Fig.~1A illustrates how a classical Szilard engine operates. A gas particle in a box is attached to a heat reservoir. After measuring its position, a movable partition is inserted, and an external load is connected. The section with the particle is then expanded adiabatically, performing work on the load. 
An analogous quantum version of the Szilard engine made of a two-level spin is presented in Fig.~1B. 
The two-level system undergoes a heat bath and reaches thermal equilibrium. Then Maxwell's demon detects the quantum states, corresponding to the particle's position measurement. Based on the measurement results, an adiabatic transition is applied, equivalent to the particle doing work on the external load, thereby increasing the potential energy of the load. 

In our realization, a single $^{171}{\rm Yb}^+$ ion confined in the blade trap serves as the carrier of the information engine. The two internal states $\ket{F=0,m_F=0}$ and $\ket{F=1,m_F=0}$ on the $^2 S_{1/2}$ level of the $^{171}{\rm Yb}^+$ ion is represented for $\ket{\uparrow}$ and $\ket{\downarrow}$, respectively, where the energy gap is $\omega_0=(2\pi)12.6428~\rm{GHz}$. The vibrational mode along the x-axis with the frequency of $\omega_{\rm x}=(2\pi)3.1~\rm{MHz}$ is used for the mechanical oscillator as a quantum battery. Thus, the Hamiltonian of the system that consists of a qubit and a mechanical oscillator is written as $H_{\mathrm{sys}}=\frac{1}{2}\hbar\omega_0\sigma_{\rm z}+\hbar\omega_{\rm x}a^{\dag}a$. In the Szilard engine, the work from the particle is entirely converted into an increase in the potential energy of the external load. Similarly, we use the rotating frame $U(t)=e^{-\frac{i}{2}(\omega_0-\omega_{\rm x})\sigma_{\rm z}t}$ to make the energy spacings of the internal state and the oscillator equal to $\hbar\omega_{\rm x}$, ensuring energy conservation for the two-level system and the quantum battery during work generation (Fig.~1B).

Our engine absorbs heat from a heat reservoir and cyclically converts information into work. As illustrated in Fig.~2A, the heat reservoir is simulated by applying a microwave field and a 369.5 nm laser beam, which, when in contact with the system, establishes a thermal state corresponding to the set temperature regardless of the system's initial state. We first prepare the internal state in the $\ket{\downarrow}$ state using the standard optical pumping method \cite{olmschenk2007manipulation}. By applying the microwave pulse, we create a superposition of the $\ket{\downarrow}$ and $^2S_{1/2}\ket{F=1,m_F=1}$ states, and then we pump the $^2S_{1/2}\ket{F=1,m_F=1}$ state to the $\ket{\uparrow}$ state using a $\sigma_+$ and $\sigma_-$ polarized laser beam. After these procedures, a mixed state of $\ket{\uparrow }$ and  $\ket{\downarrow}$ is prepared, which is a thermal equilibrium state \cite{linden2009quantum} with the temperature $k_{\rm B}T=\hbar \omega_{\rm x}\left[{\rm ln} \frac{1-P_\uparrow}{P_\uparrow}\right]^{-1} $. Here, the temperature of the bath is adjusted by controlling the duration of the microwave pulse. 
The results of the quantum state tomography confirm that the simulated heat bath generates a thermal state with the temperature as our expectation (see supplementary).

The internal state measurement, analogous to the particle position measurement in the classical Szilard engine, is performed through fluorescence detection. As shown in Fig.~2B, the $\ket{\uparrow}$ state produces scattering photons from the laser beam \cite{olmschenk2007manipulation}, which can affect the mechanical oscillator due to photon recoil. To mitigate this effect, we develop a high-efficiency imaging system with a 0.6 numerical aperture (NA) objective lens, capable of collecting about 10\% of the scattered photons. This enhanced collection efficiency enables us to reduce the detection duration to 20 $\mu$s, achieving a state-discrimination fidelity of $98.7\%$. During the internal state detection, the average phonon number increases by 0.5 for the mechanical oscillator's ground state, corresponding to around 235 times the photon absorption and emissions. A detailed analysis of the measurement-induced disturbance is provided in the supplementary material.

Work generation and transfer to a quantum battery are performed through the adiabatic transition that couples the internal state with the mechanical oscillator \cite{an2015experimental,um2016phonon}. Based on the measurement outcome, different feedback is applied accordingly: if the ion is detected in the $\ket{\downarrow}$ state, no further action is taken. However, if the ion is found in the $\ket{\uparrow}$ state, an adiabatic transition is applied. The adiabatic transition can transfer the $\ket{\uparrow,n}$ state to the $\ket{\downarrow,n+1}$ state for a fixed duration regardless of $n$. Here, the average generated work on the system in a cycle is given as $-\overline{W}=P_{\uparrow} \hbar \omega_{\rm x}$, which is entirely transferred to the mechanical oscillator during the ideal adiabatic transition. The actual experimental results will be discussed later in Fig.~4B (also in supplementary). The adiabatic transition is accomplished by applying two counter-propagating Raman laser beams as the Rabi frequency $\Omega$ and the frequency detuning of the Raman beams $\delta$ are controlled as shown in Fig.~2C. Specifically, both parameters are modulated as $\Omega(t)=\Omega_0 \sin(\frac{\pi t}{\tau})$ and $\delta(t) = \delta_0 \cos(\frac{\pi t}{\tau}) $ \cite{an2015experimental,um2016phonon}, where, $\Omega_0=(2\pi)21$ kHz, $\delta_0= 3\Omega_0$ and $\tau =244 \mu$s. Finally, the stored energy in the mechanical oscillator can be read out by applying phonon-number-resolving detection \cite{an2015experimental,um2016phonon,shen2018quantum} (see also supplementary text). 

Fig.~3A illustrates the increase in the average phonon number of the mechanical oscillator, representing the quantum battery's average energy, over successive cycles of the quantum Szilard engine. Prior to engine operation, the mechanical oscillator is cooled to near its ground state, with an initial average phonon number of 0.009(2). We execute the engine for up to 10 cycles, which is in contact with a heat bath of various temperatures. Then, we perform phonon-number-resolving detection to determine the distribution of phonon-number states, as exemplified in Fig.~3B. The average phonon numbers in our experiments consistently increase with the number of cycles, as expected theoretically. However, the results are higher than the ideal values overall. 

This discrepancy primarily stems from three factors: the disturbance of the mechanical oscillator by fluorescence detection, imperfections in phonon-resolving detection, and the heating of the mechanical oscillator. Among these, heating induced by electric noise is minimal, with a heating rate of 0.045(7) phonons per millisecond, contributing approximately 0.02 phonons per cycle. We quantify the phonon-number-resolving detection error experimentally (see supplementary materials) and characterize the photon disturbance impact using displacement operators \cite{de2022error,hempel2013entanglement}. Our simulations indicate that the fluorescence detection error can be effectively modeled by considering the disturbance from approximately 235 absorbed and emitted photon pairs (see supplementary materials). Dashed lines in Fig.~3A present the theoretical values after incorporating these errors, demonstrating a strong agreement with the experimental results. This suggests that the main sources of error are disturbances caused by fluorescence photons and imperfections in phonon-resolved detection.

It is important to note that not all energy stored in a quantum battery is extractable. For example, while heating can increase the average energy of the mechanical oscillator, this increase does not contribute to useful energy. The maximum extractable work from a quantum system is quantified by ergotropy \cite{allahverdyan2004maximal}, which serves as a measure of the useful energy in a quantum battery. Ergotropy is defined as:
\begin{equation}
\mathcal{E} = {\rm Tr}[H_{\rm QB}(\rho-\sigma)],
\end{equation}
where $H_{\rm QB}$ represents the Hamiltonian of the quantum battery system, $\rho$ denotes the density matrix, and $\sigma$ is derived by rearranging $\rho$ in descending order of its eigenvalues \cite{francica2020quantum}. This formula is applicable to finite quantum systems. In our scheme, we set the cutoff of the mechanical oscillator at a ten-phonon state, as the probability of finding higher phonon number states is negligible. Fig.~4A illustrates the steady increase in the ergotropy of our quantum battery as the engine operates. Notably, the useful energy charged into the quantum battery is significantly lower than the average energy increase depicted in Fig.~3A. For the highest temperature reservoir in our experiment ($P_{\uparrow}\approx 0.5$), the average energy after 10 cycles reaches 5 $\hbar \omega_{\rm x}$.  In contrast, the ergotropy in the ideal case would be 2.92 $\hbar \omega_{\rm x}$, while our experimental results yield an ergotropy of only 0.85 $\hbar \omega_{\rm x}$.  The primary factor contributing to this deviation from the ideal ergotropy is the presence of disturbance by scattering photons generated during fluorescence detection. The disturbance introduces additional energy into the system, which, while increasing the average energy, does not contribute to the useful, extractable energy of the quantum battery.

We then investigated the information-to-energy conversion efficiencies of our quantum information engine.
The second law of thermodynamics states that a system starting from thermal equilibrium has $\langle\Delta F-W\rangle\le 0$, where $\Delta F$ represents the free energy change of the system and $W$ denotes the work done on the system. This implies that in each cycle of the engine, the work extracted from the system should not exceed the free energy change. However, measuring the internal state reduces the Shannon entropy of the system by $k_{\rm B}I$ and consequently increases the free energy, allowing more work to be extracted \cite{parrondo2001szilard}. Here, $I=-{P_\uparrow}\ln{P_\uparrow}-(1-{P_\uparrow})\ln{(1-P_\uparrow)}$ is the mutual information content between the internal state and the measurement outcome \cite{shannon1948mathematical}. Taking into account the impact of information, the new inequality becomes $\langle\Delta F-W\rangle\le k_{\rm B} TI$.
As a result, the information becomes the source of work done by the engine and change in free energy, with the maximum value of the sum of these two components being $k_{\rm B} TI$. 

To comprehensively evaluate our quantum information engine and quantum battery system, we calculated two different efficiencies: the information-to-work conversion efficiency, denoted as $-\frac{\overline{W}}{k_{\rm B}T I}$, and the charging efficiency, represented as $\frac{\overline{\mathcal{E}}}{k_{\rm B}T I}$. In these expressions, $-\overline{W}$ and $\overline{\mathcal{E}}$ signify the average work and ergotropy produced by the engine per cycle, respectively. 
Here, the work done by the engine can be determined by measuring the internal state before and after the feedback control (see supplementary), as it should be equal to the decrease in the energy of ion's internal state. Especially, when $P_{\uparrow}\approx 0.5$, the engine reaches its maximum power of $0.0011\hbar \omega_x$ per microsecond.
In an ideal scenario, all the work done by the engine is converted into energy stored in the quantum battery. However, due to the heating mainly caused by photon disturbance, the energy stored in the battery will exceed the work done by the engine.
Nevertheless, while the photon disturbance increases the energy stored in the battery, it reduces the battery's ergotropy, as shown in Fig.~4A.
It is observed that as the temperature increases, the efficiency drops from nearly 100\% to nearly 0\%, referring to Fig.~4B. This is because the rate at which mutual information increases with temperature is much faster than the work done by the engine.
When the number of engine cycles is sufficiently high, the ergotropy approaches the work done by the engine (see supplementary), and the charging efficiency almost coincides with the information-to-work conversion efficiency in the ideal case. However, due to the limitations of phonon-number-resolving detection for high phonon number states, our demonstration was restricted to only 10 cycles of engine operation. In this context, the ergotropy is significantly smaller than the work done by the engine, leading to a decrease in charging efficiency. Especially at low temperatures, the ergotropy shows minimal increase with the engine's operation.
In the experiment, the maximum achievable charging efficiency is 4.60\%, which is 51.2\% of the ideal scenario.

Our experimental single-ion information engine, implemented at the fundamental quantum scale, can serve as a platform for further experimental development, including the practical application of mechanical energy. The integration of quantum memory replacing its classical counterpart would enable the exploration of engines that surpass the efficiency of classical information engines by leveraging quantum information resources like coherence and entanglement \cite{scully2003extracting,kim2011quantum,park2013heat,zurek2018eliminating,myers2022quantum}. Additionally, our work can pave the way for realizing devices capable of reaching quantum advantage in battery charging \cite{campaioli2024colloquium,mitchison2021charging,yan2023charging,ferraro2018high-power,gyhm2022quantum}. Furthermore, the fast detection technique we've introduced for surpressing measurement disturbance on mechanical oscillators may find applications in other areas of quantum information science, which includes improving quantum error correction for bosonic qubits \cite{de2022error} and exploring multi-mode number-resolving detection \cite{chen2023scalable}.

\bibliography{scibib}

\bibliographystyle{Science}

\vspace{5pt}
{\noindent\Large\textbf{Acknowledgments\vspace{10pt}}\\} 
This work was supported by the Innovation Program for Quantum Science and Technology under Grants No. 2021ZD0301602, and the National Natural Science Foundation of China under Grants No.92065205, No.11974200, No.62335013, and No.12304551.

\clearpage
\begin{figure}[t]
    \centering
    \hspace{-0.4cm}
    \includegraphics[width=0.8\textwidth]{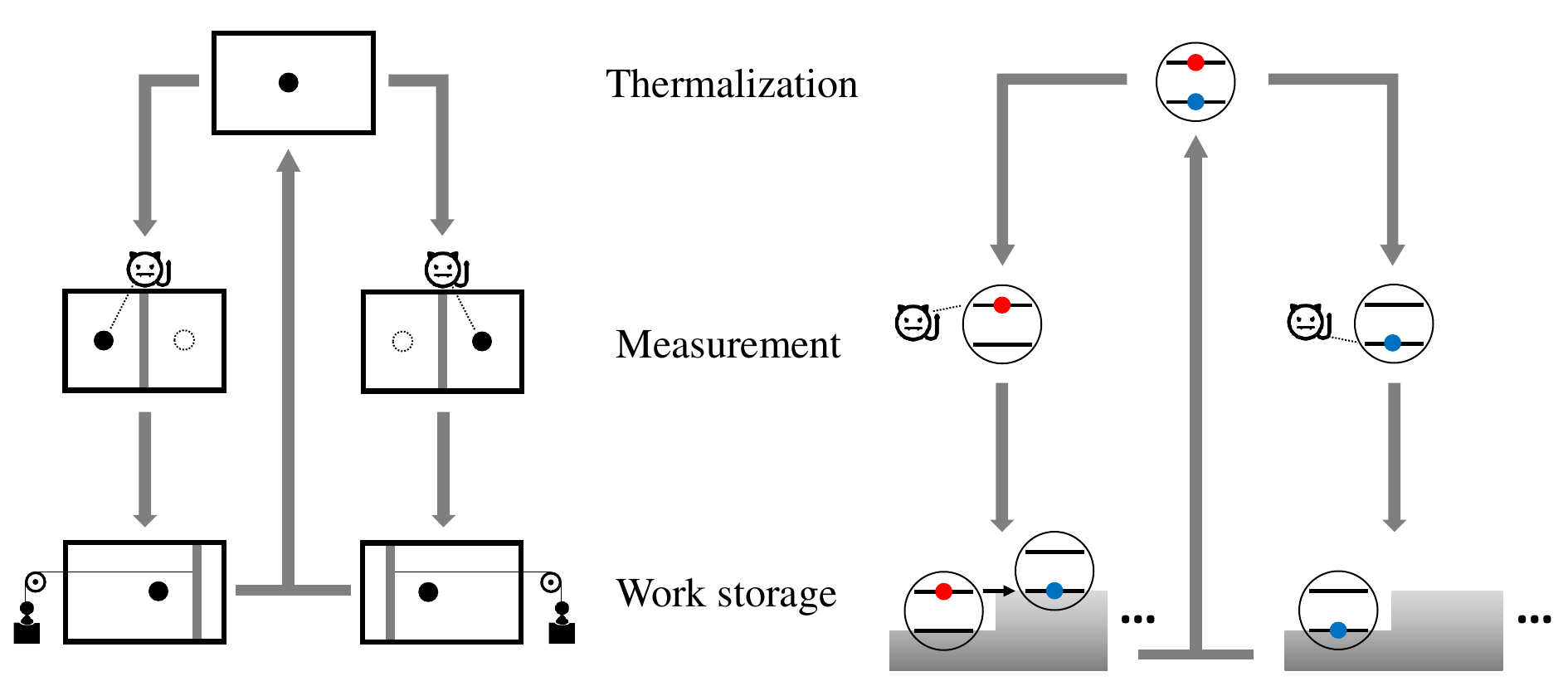}
    \put(-350,155){\textbf{A}}
    \put(-140,155){\textbf{B}}
    \vspace{-5pt}
\end{figure}

\noindent Figure 1: {\bf Schematic diagrams of the classical and quantum Szilard engine.} (A) The originally proposed Szilard engine consists of a single particle in a box that can be in contact with a heat reservoir. Work is generated by applying a load on the same side as the particle's position, which is cyclically operated.    
(B) Analogous to the classical Szilard engine, the quantum information engine consists of a two-level spin that can be in contact with a heat bath. The engine does work and stores it in a quantum battery when the $\ket{\uparrow}$ state is detected, and the process is cyclically operated. Here, the height of the steps represents the energy stored in the quantum battery.

\clearpage
\begin{figure}[t]
    \centering    
    \includegraphics[width=1\textwidth]{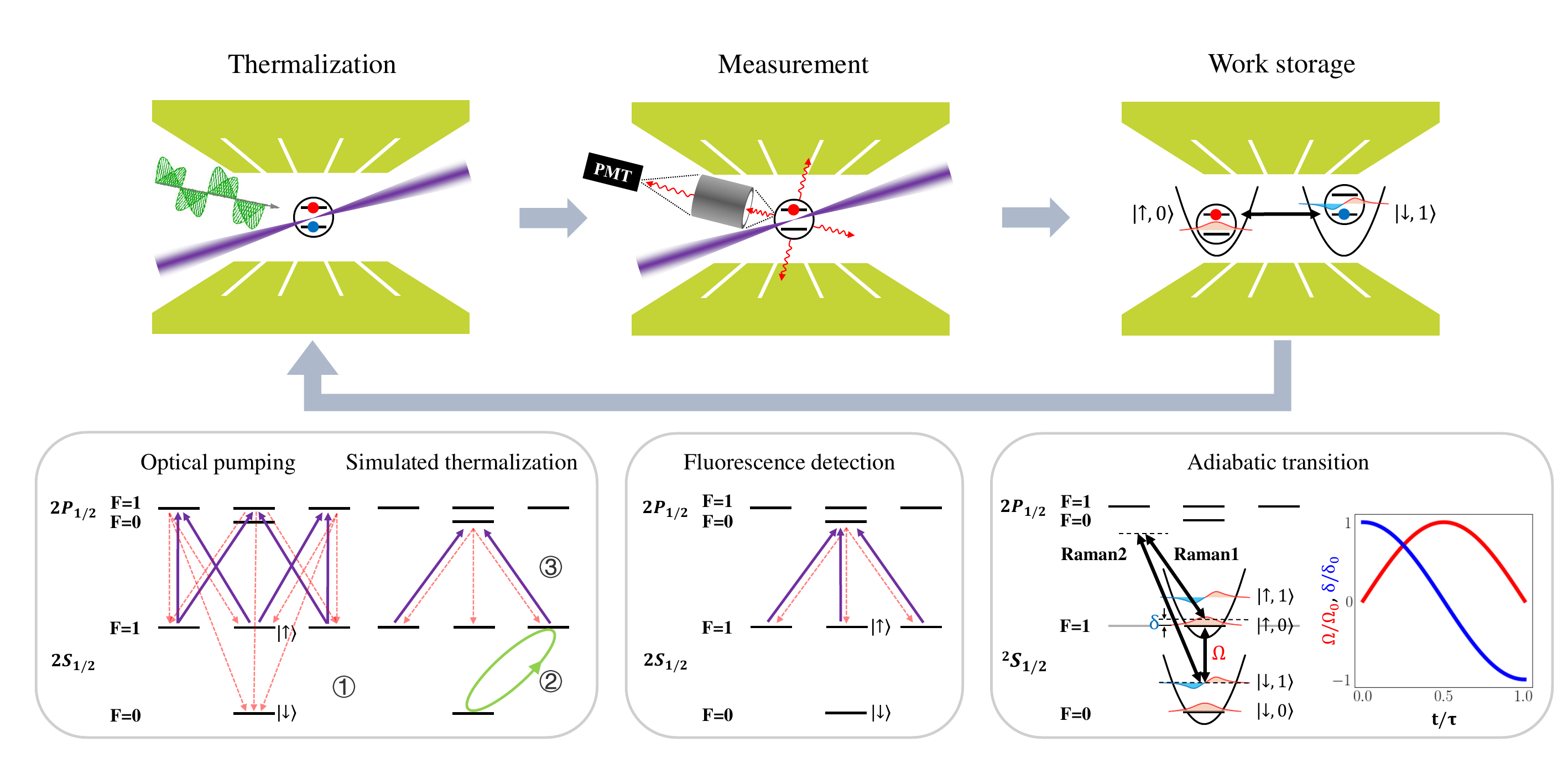}
    \put(-412,206){\textbf{A}}
    \put(-273,206){\textbf{B}}
    \put(-130,206){\textbf{C}}
\end{figure}
\noindent Figure 2: {\bf Experimental implementation of the quantum Szilard engine.} (A) The heat bath to the \Yb ion is realized by applying microwave pulses and 369.5 nm laser beams, represented by green and purple colors, respectively. The procedure and the effect of applying them are shown below in the energy-level diagram of the ion. The circled number indicates the steps of setting population distribution on the qubit states, corresponding to the temperature of the heat bath. Here, dashed lines indicate the scattering process of the \Yb ion. 
(B) The qubit state detection is performed by applying a 369.5 nm laser and collecting the scattering photons when the ion is in the $\ket{\uparrow}$ state. The photon counts are read out through a photomultiplier tube. Photon scattering disturbance is minimized by collecting photons using 0.6 NA objective lens, which helps in reducing the measurement duration.  (C) Work is generated and stored from an adiabatic transition if the state is detected as $\ket{\uparrow}$. The beat-note from two counter-propagating Raman beams couples the $\ket{\uparrow,n}$ and $\ket{\downarrow,n+1}$ states. The Rabi frequency $\Omega$ and the frequency difference $\delta$ are temporally modulated, as shown by the red and blue lines.

\clearpage
\begin{figure}[t]
    \centering
    \begin{subfigure}[c]{0.8\textwidth}
        \includegraphics[width=1.\textwidth]{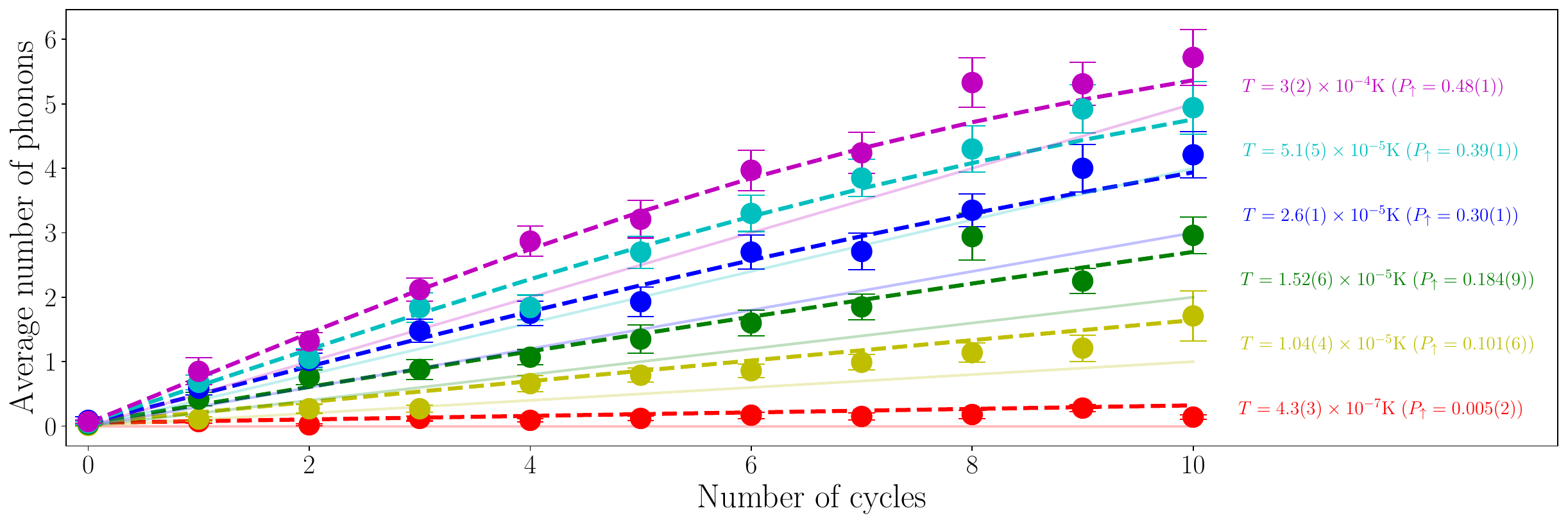}
        \put(-360,125){\textbf{A}}
        \begin{tikzpicture}[overlay, remember picture]
            \draw[gray, line width = 1.5pt] (-8.65cm, 0.7cm) rectangle (-8.27cm, 2.53cm);
            \draw[gray,->, line width = 2pt] (-8.27cm, 0.7cm) -- (-7.5,-0.32);
        \end{tikzpicture}
    \end{subfigure}
    \vspace{20pt} 
    \begin{subfigure}[c]{0.8\textwidth}
        \includegraphics[width=1\textwidth]{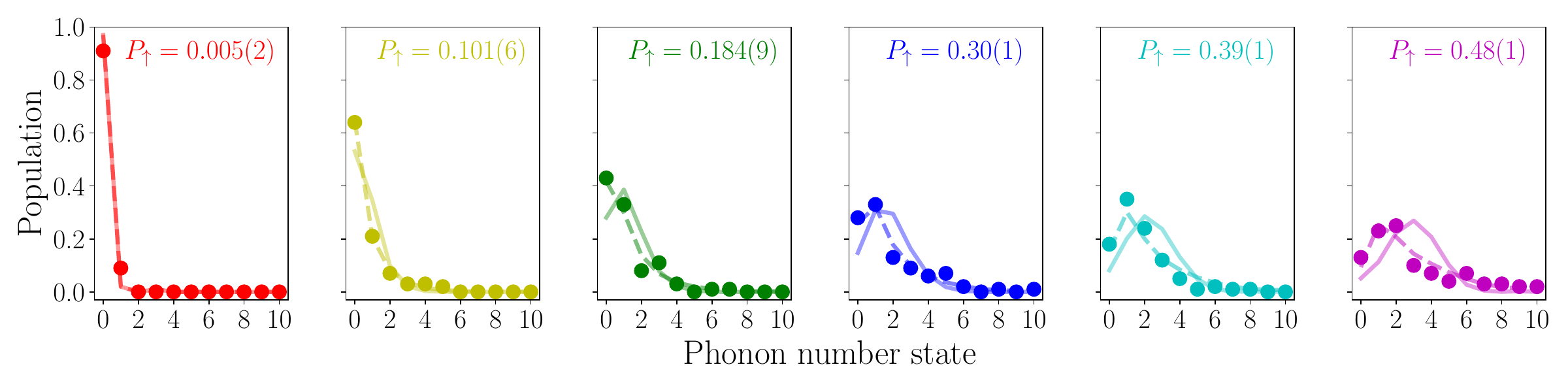}
        \put(-360,90){\textbf{B}}
    \end{subfigure}
    \vspace{-20pt}
\end{figure}
\noindent Figure 3: {\bf The quantum battery charged by the information engine at different temperatures.} (A) The average phonon numbers of the mechanical oscillator, which are proportional to the average energy of the quantum battery, increase with the number of engine cycles. We run the engine at 6 different temperature baths, represented by different colors. Filled circles depict the experimental results, and the error bars are the standard deviation obtained from the bootstrap method. Solid lines represent the expected energy in an ideal situation, while dashed lines represent the theoretical results that include the disturbance by fluorescence detection and the imperfection of phonon-number-resolving detection.  (B) Examples of phonon number distribution are obtained by applying the number-resolving detection method after the engine runs for 4 cycles at the temperature corresponding to panel (A). The filled circles, solid lines, and dashed lines are used consistently with those in panel (A).

\clearpage
\begin{figure}[t]
    \centering
    \begin{subfigure}[c]{0.45\textwidth}
        \includegraphics[width=1\textwidth]{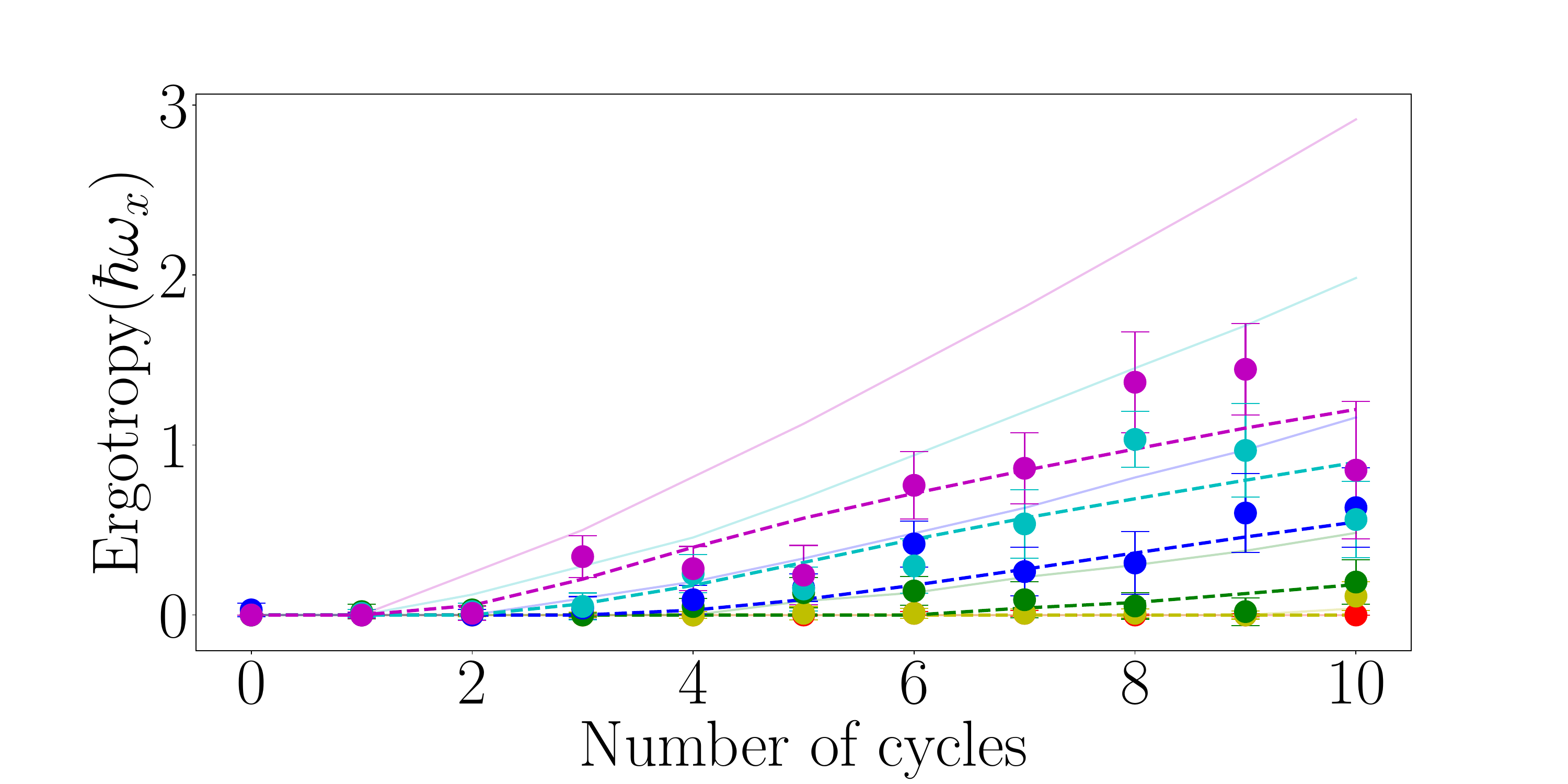}
        \put(-205,90){\textbf{A}}
    \end{subfigure}
    \\
    \begin{subfigure}[c]{0.45\textwidth}
        \includegraphics[width=1\textwidth]{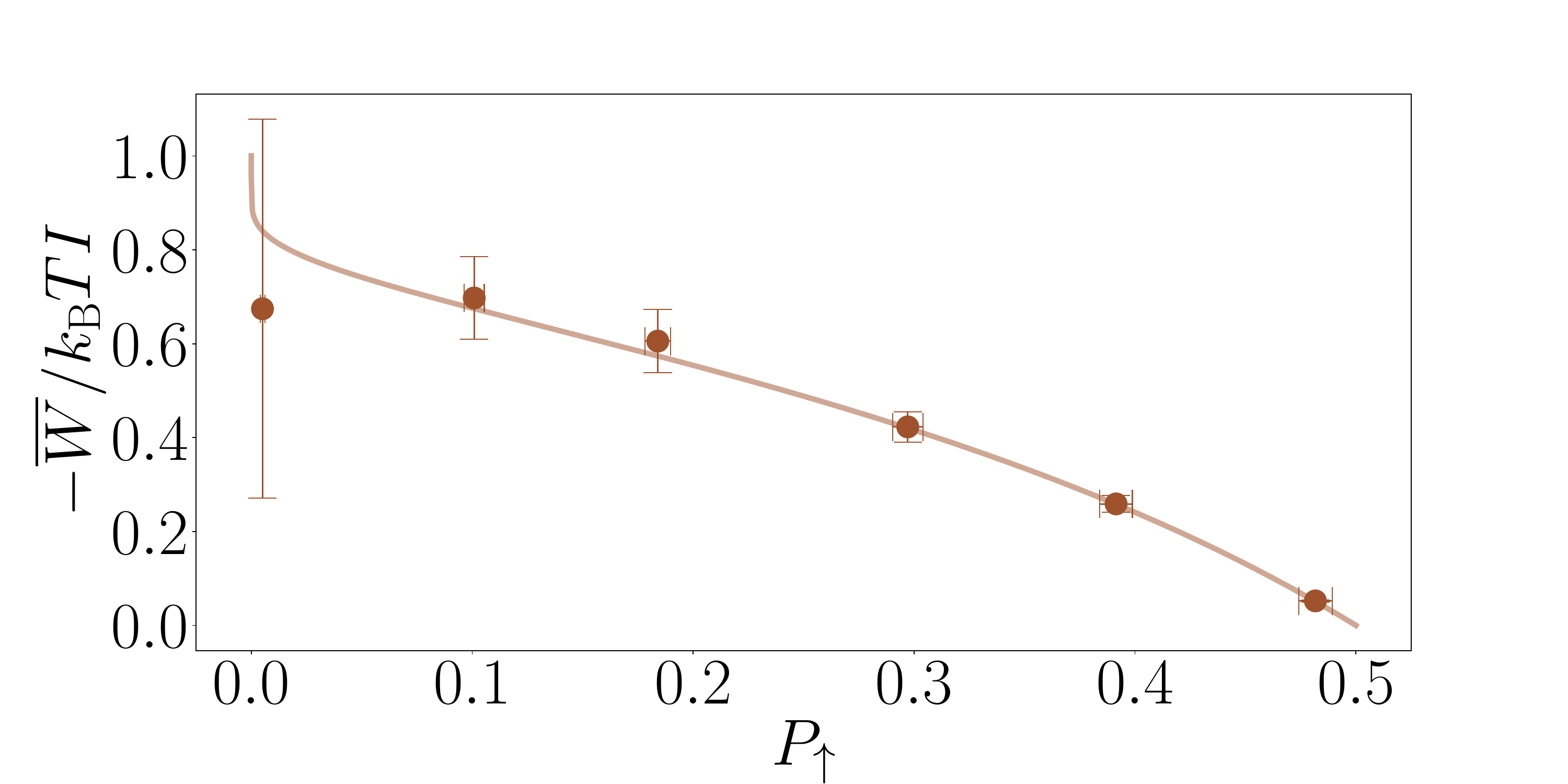}
        \put(-205,90){\textbf{B}}
    \end{subfigure}
    \\
    \begin{subfigure}[c]{0.45\textwidth}
        \includegraphics[width=1\textwidth]{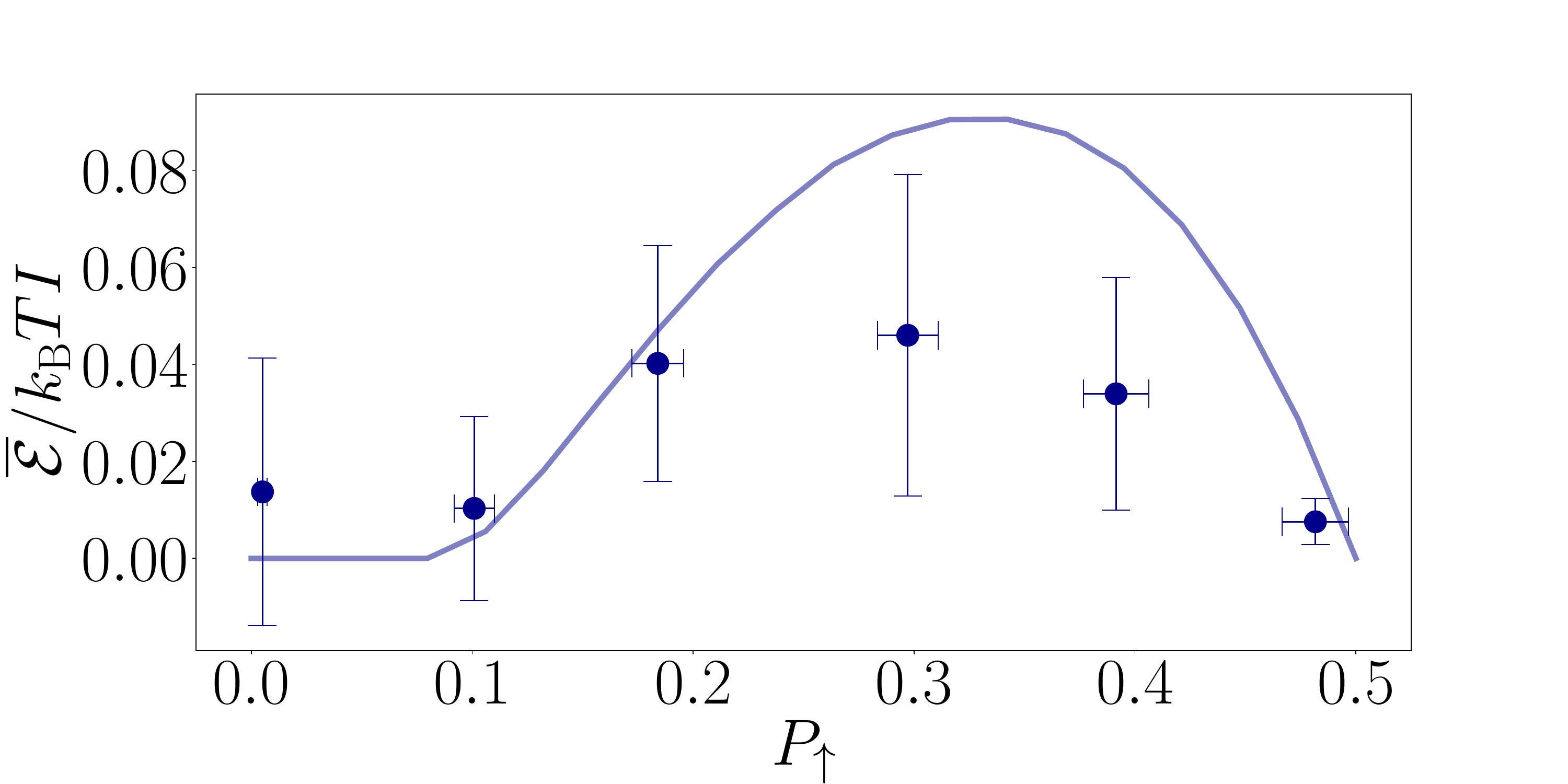}
        \put(-205,90){\textbf{C}}
    \end{subfigure}
\end{figure}
\noindent Figure 4: {\bf The ergotropy and the information-to-energy efficiencies.} (A) The ergotropy of the quantum battery with the number of cycles. Various colors represent different temperatures of the heat bath consistent with Fig.~3. The solid and dashed lines represent the ideal values and the simulation of ergotropy, respectively. The circles depict the experimental results for ergotropy, with error bars derived from the standard deviation of the bootstrap method. (B) The efficiency of information-to-energy conversion. The circles indicate the values of $-\frac{\overline{W}}{k_{\rm B}TI}$ obtained from the experiment, while the line represents the efficiency in the ideal scenario. (C) The charging efficiency. The circles, as well as the line, illustrate the experimental and ideal values of $\frac{\overline{\mathcal{E}}}{k_{\rm B}TI}$, respectively. The highest efficiency obtained in the experiment is 4.6\%, achieved at $P_{\uparrow}=0.30(1)$.

\clearpage
{\noindent\Large\textbf{Supplementary text}\\ \\}
\noindent\textbf{Heat bath} \quad The heat bath is simulated using a combination of microwave and 369.5 nm laser. By adjusting the duration of the microwave and laser pulses, the population of the $\ket{\uparrow}$ state after the heat bath, denoted as $P_{\uparrow}$, is prepared to be 0, 0.1, 0.2, 0.3, 0.4, 0.5. Subsequently, through quantum state tomography, the density matrix of the internal state of the ion is obtained, as illustrated in Fig.~S1. The off-diagonal elements of the density matrices are nearly zero, indicating that the ion is in the mixed states, which aligns with our expectations.\\

\noindent\textbf{Adiabatic transition} \quad A successful $\pi$-pulse adiabatic transition transforms the state $\ket{\uparrow,n}$ to $\ket{\downarrow,n+1}$, and its duration should not depend on the initial phonon number state. To test this, we first prepare the phonon number state from $\ket{0}$ to $\ket{10}$ with the internal state in $\ket{\uparrow}$. Then we perform a $\pi$-pulse adiabatic transition and monitor the evolution of the internal state over time. As illustrated in Fig.~S2, the ion initially in different phonon number states transitions to the $\ket{\downarrow}$ state after the same duration, indicating that our adiabatic transition is almost perfect.\\

\noindent\textbf{Fast fluorescence detection} \quad In our experiment, we use an objective lens with a numerical aperture (NA) of 0.6 in the imaging system. Approximately 10\% of all the scattering photons can be collected by this high-NA lens. We reduced the detection time interval to 20 microseconds, and the detection results for the dark state $\ket{\downarrow}$ and the bright state $\ket{\uparrow}$ are illustrated in Fig.~S3. We improve detection fidelity by correcting the aberrations of the imaging system and employing multiple optical filters. The fidelity of the dark state is 97.8\%, and for the bright state, it is 99.6\%, with an average fidelity of 98.7\%. During fast fluorescence detection, we recorded an average photon count of 7.3 from the bright state $\ket{\uparrow}$. Considering the 10\% collection efficiency of the objective lens and the 40\% PMT efficiency, the total count of scattering photons is approximately 183. Taking into account photon losses in propagation and filtering, the actual number of scattering photons is even higher.\\

\noindent\textbf{Phonon-number-resolving detection} \quad The phonon-number-resolving detection enables the determination of the distribution of phonon number states, as demonstrated in Fig.~S4. After applying optical pumping, adiabatic transition, and carrier transition sequentially, the ion initially in the $\ket{0}$ state is transitioned to the bright state $\ket{\uparrow}$, which can be distinguished by fluorescence detection. Conversely, the ion in other phonon number states will reduce by 1 phonon and will be in the dark state $\ket{\downarrow}$. This process is reiterated until the bright state is observed, with the total number of repetitions, n+1, serving as an indicator that the ion occupies the $\ket{n}$ state.\\

\noindent\textbf{Error anlysis} \quad The impact of imperfections in phonon-number-resolving detection, as well as disturbances caused by fluorescence photons, is quantified in our experiment. Here, we use a column vector to represent the population distribution of the ion's phonon number states. The vector $(a_0,a_1,...a_{10})^T$ indicates the population of the phonon number states from $\ket{0}$ to $\ket{10}$ as $a_0$, $a_1$, ..., $a_{10}$ respectively. A cut-off is defined on the $\ket{10}$ state since higher phonon number states are rarely occupied.
In the experiment referring to Fig.~S5, the errors arising from phonon-number-resolving detection and photon disturbance are described by matrices $M_1$ and $M_2$, respectively, through the formula:
$$ \rho_i^1=M_1\ket{i}_p \text{ and } \rho_i^2=M_1 M_2 \ket{i}_p.$$
Where $\ket{i}_p$ is the population vector corresponding to the phonon number state $\ket{i}$. It is easy to find that $M_1 = (\rho_1^1, \rho_2^1, ..., \rho_{10}^1 )$, however, obtaining the value of $M_2$ is a challenge. 

The disturbance caused by the absorption and emission of a pair of photons on the mechanical oscillator can be characterized as a displacement in phase space \cite{rasmusson2024measurement}, that is
$$
\alpha_k=\frac{i e^{i \omega_x t} \hbar k}{\sqrt{2 m \omega_x \hbar}}\left[\sqrt{f_x}+\sqrt{f_{s x}^{(k)}}\right].
$$
Here, $\omega_x$ represents the frequency of the mechanical oscillator on the x-axis, while $k$ denotes the wave vector of the 369.5 nm laser. $\sqrt{f_x}$ and $\sqrt{f_{s x}^{(k)}}$ indicate the projection of the direction of the incident and emitted photons along the x-axis. In our experiment, we utilize a resonant laser for fluorescence detection, so there is no need to introduce the term caused by the Doppler effect. 
Thus, we can use a classical Monte Carlo method to simulate the photon disturbance caused by hundreds of photons. Through numerical simulation, we found that the matrix $M_2$ can be well approximated by the photon disturbance resulting from the absorption and emission of 235 pairs of photons, as illustrated in Fig.~S6. 

The total displacement in the phase space during the fluorescence detection is given by $\alpha = \sum_i^N \alpha_k$, where N represents the number of scattered photons. It is clear that $E(\alpha) = 0$. However, the population distribution of a displaced number state is expressed as
$$
p_m(n) =|\langle m|\hat{D}(\alpha)| n\rangle|^2=\frac{n!}{m!}|\alpha|^{2(m-n)} e^{-|\alpha|^2}\left[\mathcal{L}_n^{(m-n)}\left(|\alpha|^2\right)\right]^2,
$$
which depends solely on the term $|\alpha|^2$. We find that $E(|\alpha|^2)\propto N$. Therefore, reducing the number of photons produced during fluorescence detection can help reduce photon disturbance.\\

\noindent\textbf{Work done by the engine} \quad We focused on a single cycle of the engine and designed the sequence shown in Fig.~S7. In comparison with the original engine operation sequence, an additional measurement is performed on the ion's internal state in the new sequence to verify if the feedback control functions successfully. 
During engine operation, a $\pi$-pulse adiabatic transition is applied when the $\ket{\uparrow}$ state is detected, while no action is taken when the $\ket{\downarrow}$ state is detected.
The measurement outcomes $x=\ket{\uparrow}$ and $y=\ket{\downarrow}$ indicate that the feedback control is successful, resulting in the generation of work equivalent to the energy of a phonon. The outcomes $x=\ket{\downarrow}$, $y=\ket{\uparrow}$ or $x=\ket{\uparrow}$, $y=\ket{\uparrow}$ indicate that the feedback control fails, resulting in no work being produced. The measurement outcomes $x=\ket{\downarrow}$ and $y=\ket{\downarrow}$ indicate that no action is taken, and similarly, no work is produced.
By repeatedly running the sequence, we can determine the average work done by the engine in a single cycle from multiple measurement outcomes.

Furthermore, we can verify the generalized Jarzynski equality in this single run of the engine. The generalized Jarzynski equality, which is expressed as $\langle e^{(\Delta F-W)/k_{\rm B} T}\rangle=\gamma$, explicitly demonstrates the correlation between the change in free energy and the work under feedback control \cite{sagawa2010generalized,toyabe2010experimental,parrondo2015thermodynamics}.
In this experimental sequence, the heat bath ensures that the initial and final internal states are the same, resulting in the change in free energy being 0. Therefore, the left side of the equality, $\langle e^{(\Delta F-W)/k_B T}\rangle$, can be obtained from multiple measurement outcomes. The parameter $\gamma$, which is on the right side of the equality, is measured by conducting the backward sequence as shown in Fig.~S8 and quantifies the effectiveness of the feedback control. $\gamma$ represents the sum of the probabilities of all possible paths during the time reversal, that is $\gamma=P_{\rm on}(y=\ket{\downarrow},x=\ket{\uparrow})+P_{\rm off}(y=\ket{\downarrow},x=\ket{\downarrow})$, where $P_{on}$ and $P_{off}$ are measured in the backward sequence with and without adiabatic transition.
As seen in Fig.~S8, the experimental results are consistent with the theoretical values. However, the $\gamma$ is always lower than the theoretical value, indicating that the feedback control is imperfect.\\

\noindent\textbf{Ergotropy} \quad After one complete cycle of the quantum information engine, the ion in the phonon number state $\ket{n}$ has a probability $P_\uparrow$ of transitioning to the $\ket{n+1}$ state, while the probability of remaining in the $\ket{n}$ state is $(1-P_\uparrow)$. When the engine operates for N cycles, the probability of the ion being in the phonon number state $\ket{n}$ adheres to a binomial distribution, represented as $P(N,n)=C_N^n P_\uparrow^n (1-P_\uparrow)^{N-n}$. For a sufficiently large number of cycles N, the binomial distribution can be well approximated by a normal distribution, that is, $P(N,n)\approx \frac{1}{\sqrt{2\pi N P_\uparrow (1-P_\uparrow)}}e^{-\frac{(x-NP_\uparrow)^2}{2 NP_\uparrow(1-P_\uparrow)}}$. 

The ergotropy equation (1) can be decomposed by
$$
\mathcal{E}={\rm Tr}[H_{\rm QB} \rho]-{\rm Tr}[H_{\rm QB} \sigma].
$$
Here, the diagonal elements of $\rho$ represents probability distribution of the phonon number states, where the nth element is exactly $P(N,n)$.
The first term of ergotropy then becomes 
$$
\hbar\omega_{\rm x}\int_0^{\infty}\frac{1}{\sqrt{2\pi NP_\uparrow(1-P_\uparrow)}}xe^{-\frac{(x-NP_\uparrow)^2}{2 NP_\uparrow(1-P_\uparrow)}}= NP_\uparrow\hbar\omega_{\rm x}. 
$$
Similarly, the diagonal components of $\sigma$ represent the distribution of the phonon number states that is rearranged in descending order of its probability. the nth term can be approximated as: $\frac{1}{\sqrt{8\pi NP_\uparrow(1-P_\uparrow)}}e^{-\frac{x^2}{8 NP_\uparrow(1-P_\uparrow)}}$. 
Thus, the second term becomes
$$
\hbar\omega_{\rm x}\int_0^{\infty}\frac{1}{\sqrt{8\pi NP_\uparrow(1-P_\uparrow)}}xe^{-\frac{x^2}{8 NP_\uparrow(1-P_\uparrow)}}= \sqrt{2NP_\uparrow(1-P_\uparrow)/\pi}\hbar\omega_{\rm x}. 
$$
We can find that the first term represents the work done by the engine in the ideal case. When $N$ is sufficiently large, the second term can be disregarded in comparison to the first term.

\clearpage
\begin{figure}[t]
    \centering
    \includegraphics[width=0.5\textwidth]{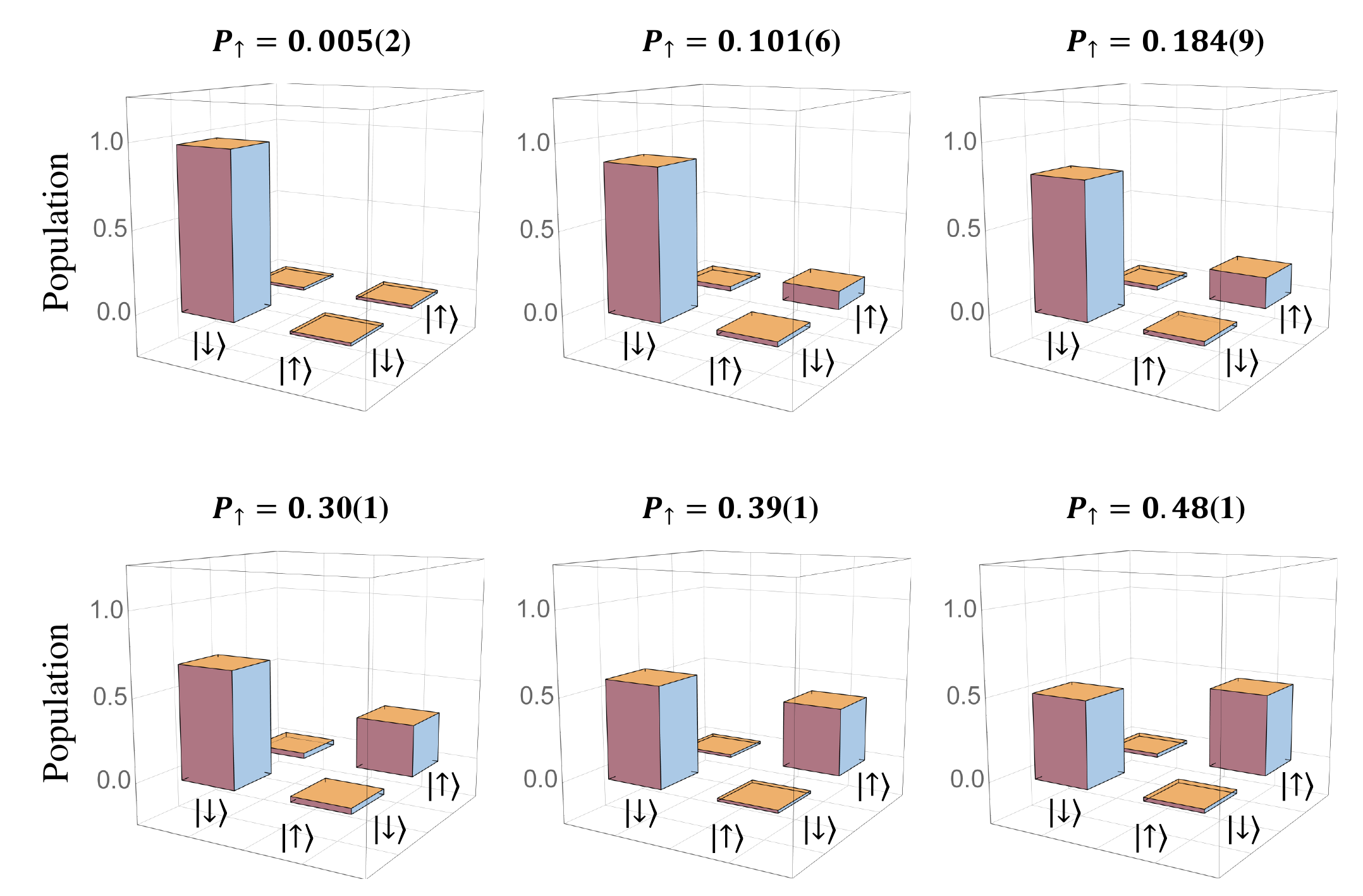}
\end{figure}
\noindent Figure S1: {\bf The density matrices of the internal state after the heat bath.} We apply heat baths of six different temperatures to the internal state and then use quantum state tomography to obtain the density matrix, as well as the measured value of $P_{\uparrow}$. The height of the bars represents the absolute value of the density matrix elements.
\clearpage

\begin{figure}[t]
    \centering
    \vspace{-20pt}
    \includegraphics[width=0.5\textwidth]{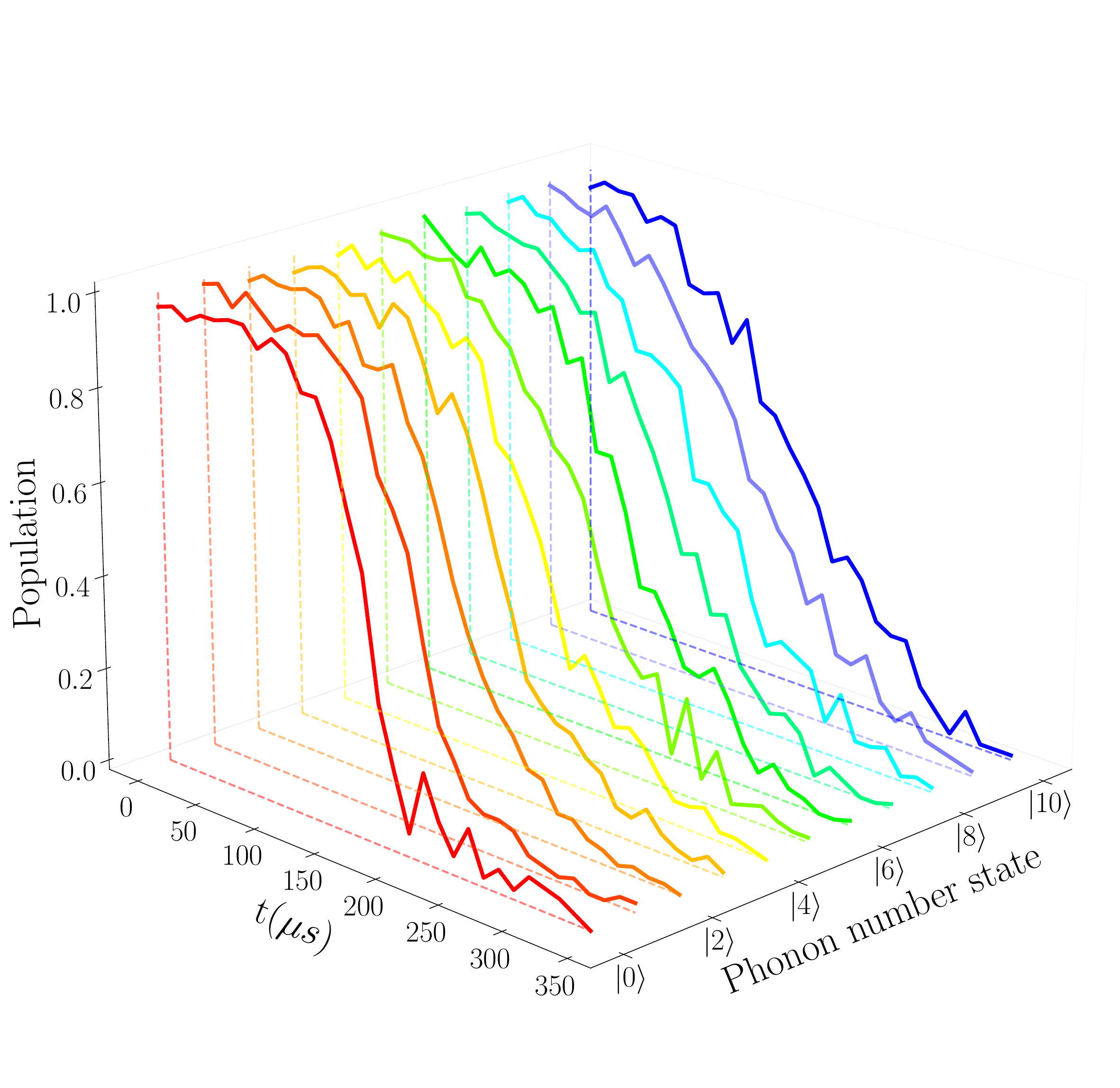}
\end{figure}
\noindent Figure S2: {\bf The evolution of the internal state under adiabatic transition.} Different colors indicate various initial phonon number states, ranging from $\ket{0}$ to $\ket{10}$. After 343 microseconds of evolution, the ion in different initial states all transitions to the $\ket{\downarrow}$ state.
\clearpage

\begin{figure}[t]
    \centering
    \includegraphics[width=0.6\textwidth]{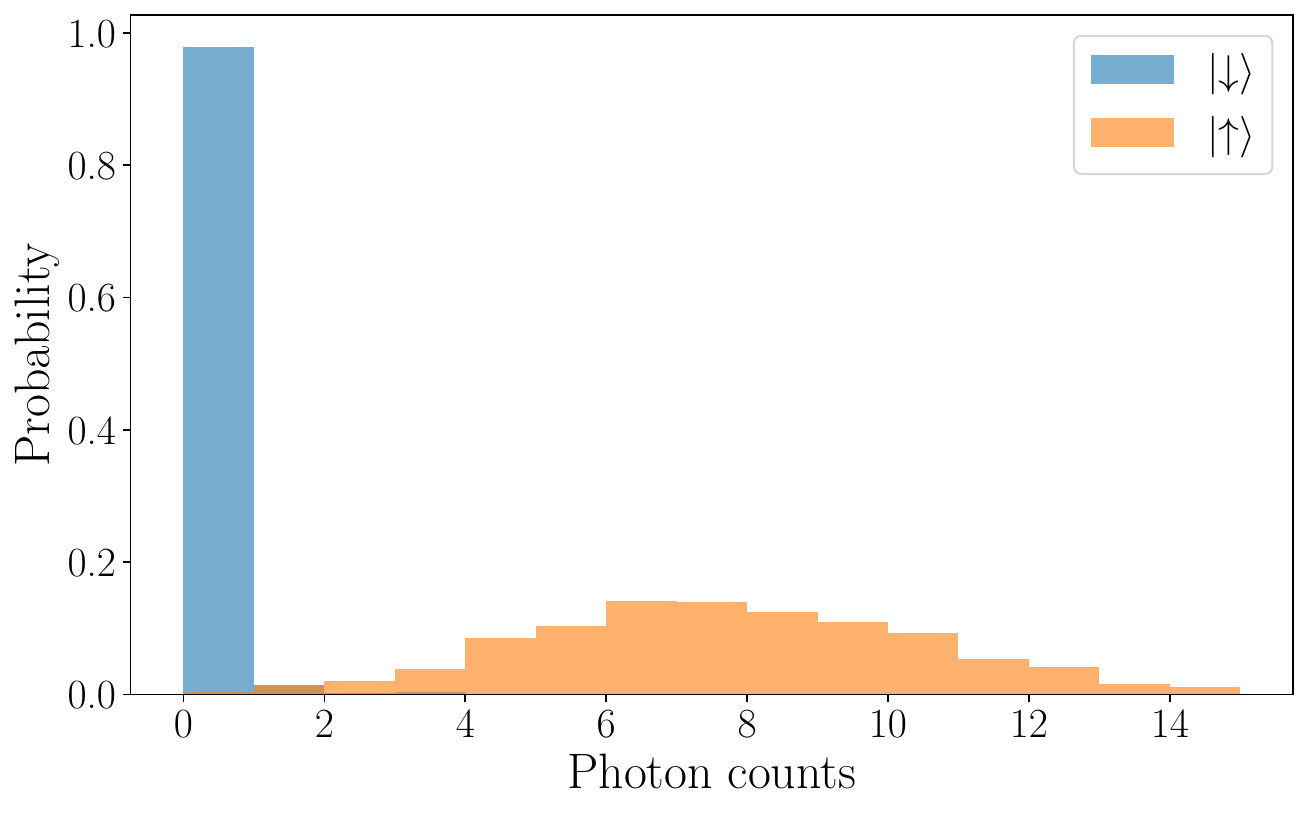}
    \begin{tikzpicture}[overlay, remember picture]
            \draw[->, line width = 2pt] (-7cm, 3cm) -- (-7.8cm,1.05cm);
    \end{tikzpicture}
    \put(-7.8cm,3.3cm){\makebox(0,0)[l]{\textbf{Threshold}}}    
\end{figure}
\noindent Figure S3: {\bf Fast fluorescence detection.} We prepared the internal state of the ion in the $\ket{\downarrow}$ and $\ket{\uparrow}$ states, respectively, and performed a 20-microsecond fast fluorescence detection. The probability distribution of the photon counts of the $\ket{\downarrow}$ and $\ket{\uparrow}$ states is shown in the figure. The threshold for distinguishing between these states is set at 1 photon count.
\clearpage

\begin{figure}[t]
    \centering
    \includegraphics[width=1\textwidth]{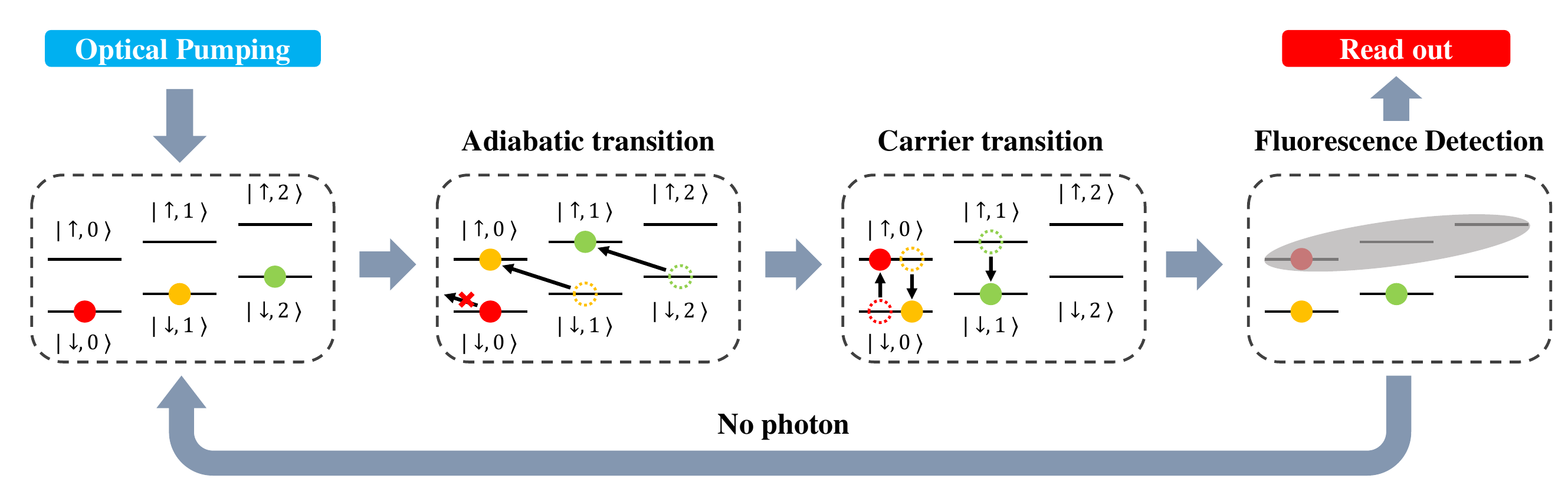}
\end{figure}
\noindent Figure S4: {\bf Sequence for phonon-number-resolving detection.} Initially, we prepare the internal state in the $\ket{\downarrow}$ state using optical pumping. We transition from the $\ket{\downarrow,n+1}$ state to the $\ket{\uparrow,n}$ state through a $\pi$-pulse adiabatic transition, except for the $\ket{\downarrow,0}$ state, which remains unchanged. Subsequently, we drive a $\pi$-pulse carrier transition using a microwave magnetic field, transitioning the $\ket{\uparrow,n}$ state to the $\ket{\downarrow,n}$ state, and the $\ket{\downarrow,0}$ state to the $\ket{\uparrow,0}$ state. Finally, we perform fluorescence detection on the ion's internal state. If a dark state $\ket{\downarrow}$ is detected, the aforementioned operation is repeated; otherwise, the sequence is terminated.
\clearpage

\begin{figure}[t]
    \centering
    \includegraphics[width=0.8\textwidth]{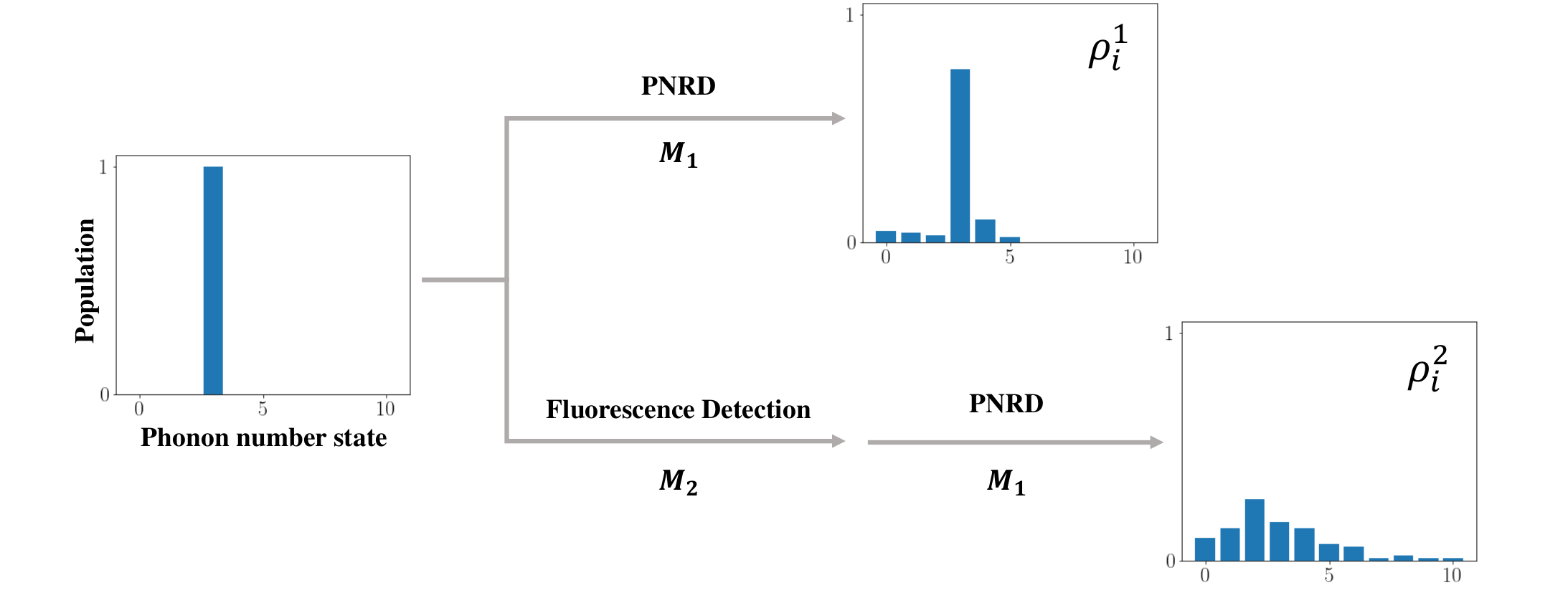}
\end{figure}
\noindent Figure S5: {\bf The sequence for quantifying the errors.} We prepare the initial state of the mechanical oscillator in the phonon number states from $\ket{0}$ to $\ket{10}$. Then we perform PNRD on these states and obtain a dataset denoted as $\rho^1=\{\rho_1^1, \rho_2^1, ..., \rho_{10}^1 \}$. The other dataset is obtained by applying additional fluorescence detection to the prepared phonon number state, followed by PNRD, and is denoted as $\rho^2=\{\rho_1^2,\rho_2^2,...,\rho_{10}^2 \}$. The error arises from the PNRD, and the fluorescence detection is represented by the matrices $M_1$ and $M_2$.
\clearpage

\begin{figure}[t]
    \centering
    \includegraphics[width=1\textwidth]{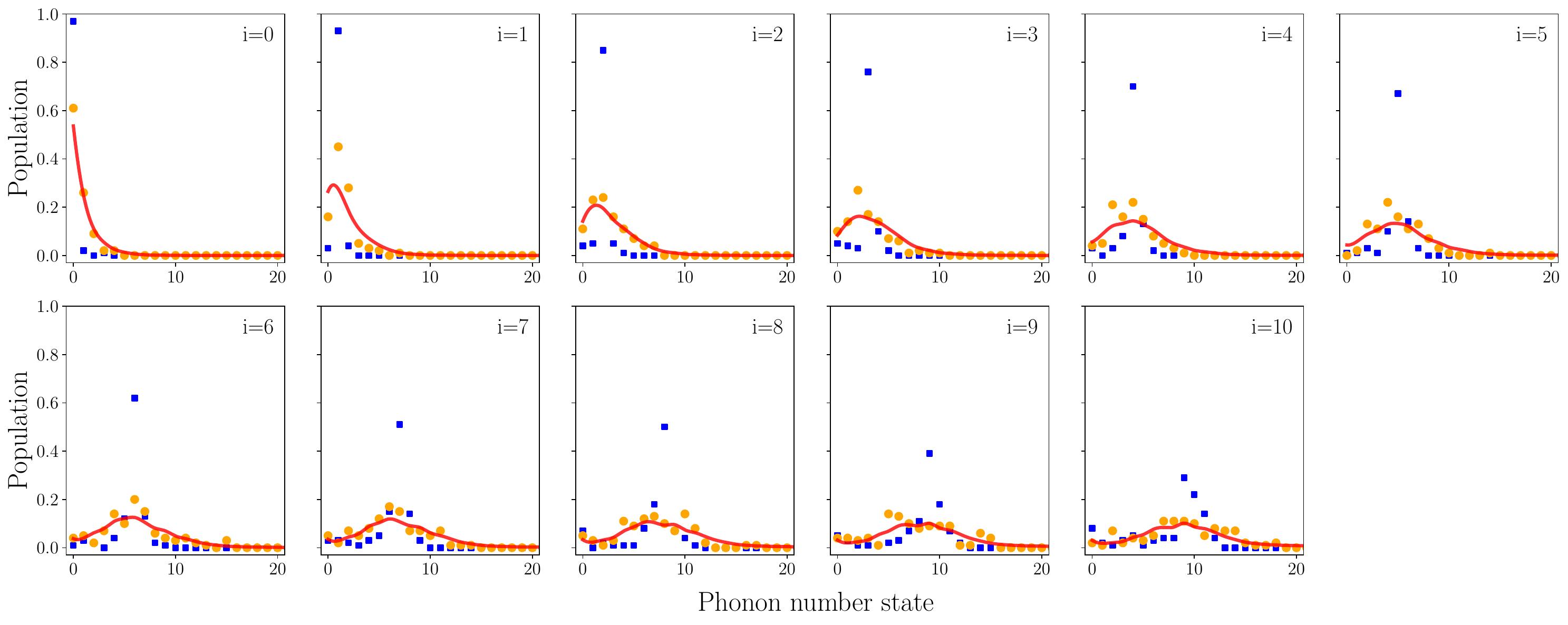}
\end{figure}
\noindent Figure S6: {\bf The measurement results of $\rho^1$ and $\rho^2$.} The blue symbols indicate the experimental results for $\rho^1$, while the orange symbols depict the results for $\rho^2$. The red lines represent the theoretical simulation results of applying the M1 and M2 (obtained from a classical Monte Carlo method) to the phonon number states.
\clearpage

\begin{figure}[t]
    \centering
    \includegraphics[width=0.5\textwidth]{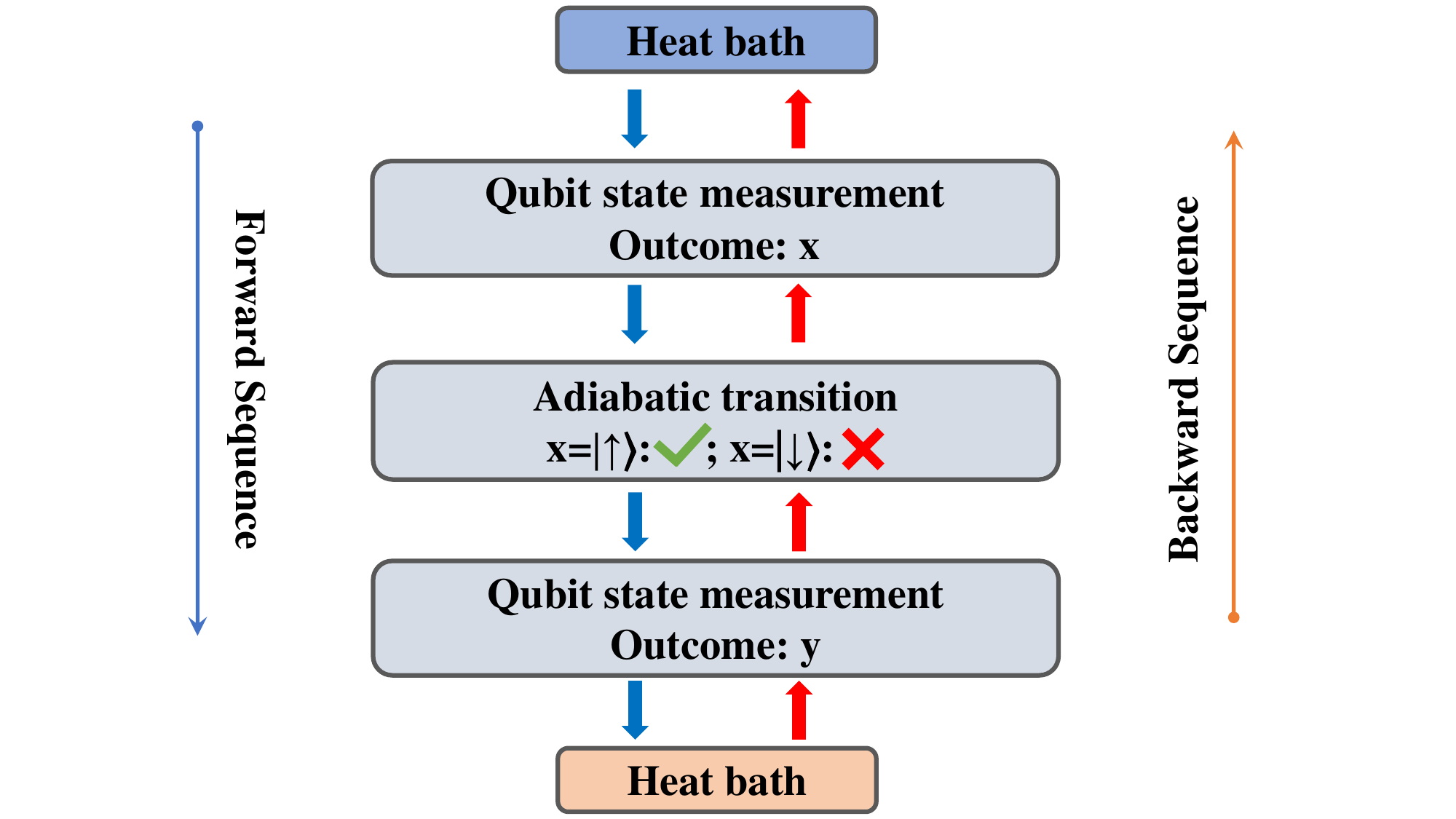}
\end{figure}
\noindent Figure S7: {\bf The sequence for one cycle of the engine.} Compared with the previous sequence, an additional measurement is applied after the feedback control to verify the success of the feedback control. In order to verify the generalized Jarzynski equality, we run the sequence in reverse order to obtain the effectiveness $\gamma$ of the feedback control.
\clearpage

\begin{figure}[t]
    \centering
    \includegraphics[width=0.5\textwidth]{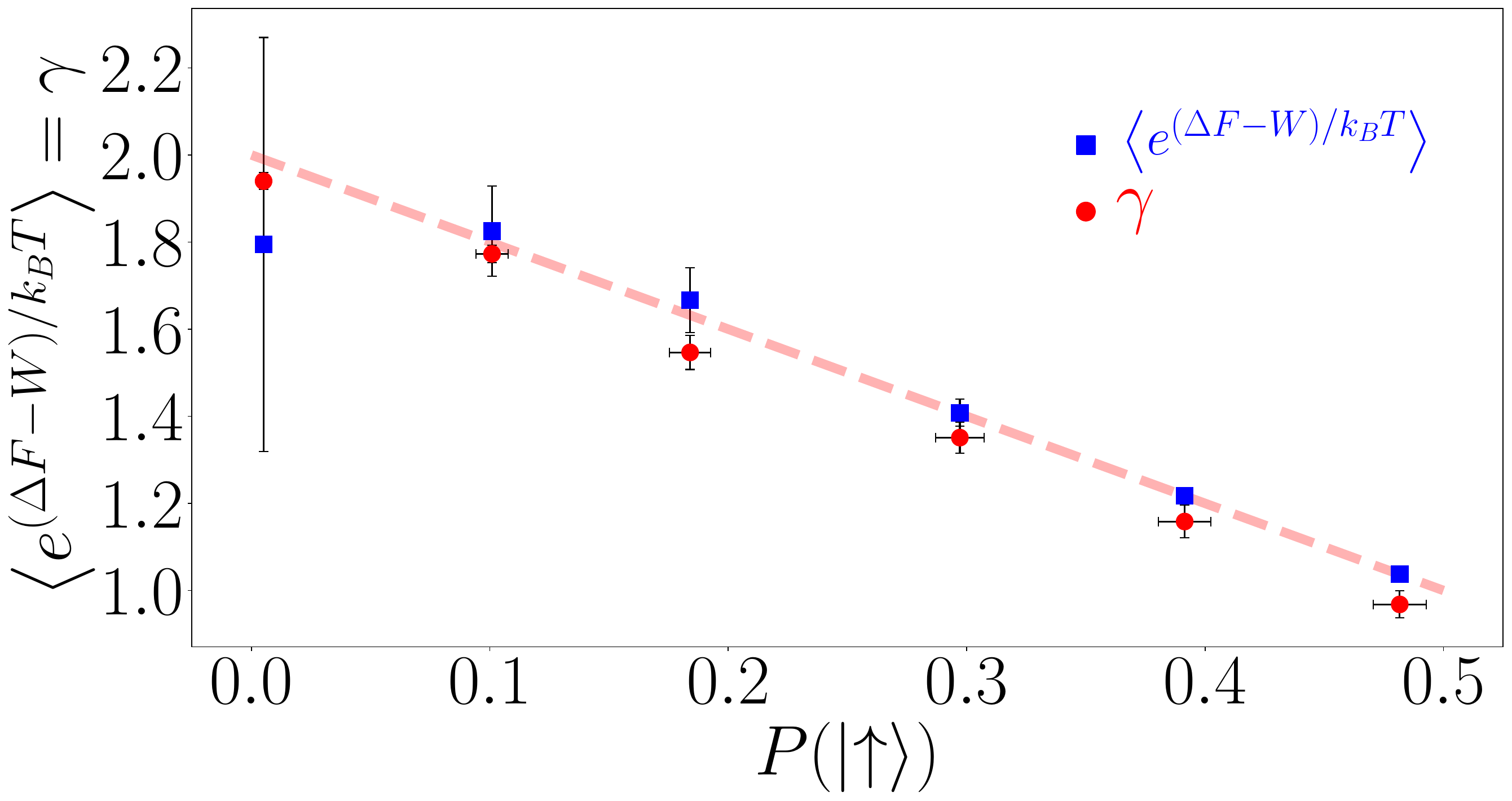}
\end{figure}
\noindent Figure S8: {\bf Verification of the generalized Jarzynski equality with varying temperatures.} Red and blue symbols represent the experimental values on the left and right sides of the equality, respectively, while the dashed line represents the ideal value.

\end{document}